\documentstyle[preprint,eqsecnum,aps]{revtex}
\begin{document}
\draft
\title{Green's Function of the Relativistic Coulomb System via Duru-Kleinert Method}
\author{De-Hone Lin \thanks{%
e-mail:d793314@phys.nthu.edu.tw}}
\address{Department of Physics, National Tsing Hua University \\
Hsinchu 30043, Taiwan}
\date{\today}
\maketitle
\begin{abstract}
In this paper the fixed-energy amplitude (Green's function) of the
relativistic Coulomb system is solved by Duru-Kleinert (DK) method. In the
course of the calculations we observe an equivalence between the
relativistic Coulomb system and a radial oscillator.
\end{abstract}
\newpage \tolerance=10000
\section{INTRODUCTION}
Undoubtedly, the hydrogen atom is one of the most interesting system in
quantum mechanics. In the beginnings of this century, it once symbolized the
success of Bohr's quantum theory and Schr\"{o}dinger's new wave mechanics,
and later pushing the progress of relativistic quantum mechanics with the
fine-structure constant and quantum electrodynamics with the observation of
the Lamb-shift.

In the past 70 years, for solving the hydrogen atom many different ways such
as the local dynamics approaches [1-5], the global dynamics approaches
[5-7], and the symmetry viewpoints [8-10] have been developed. They provide
us the diverse viewpoints of this system and stimulate the development of
some new domains (e.g. \cite{11}). Indeed, as saying by J. J. Thomson
``because a mechanical model is richer in implications than the
considerations for which it was advanced, it can suggest new directions of
research that may lead to important discoveries'' the different approach
frequently offer more abundant content than we expected.

In this paper, we apply the DK-method to the relativistic path integral and
solve the fixed-energy amplitude (Green's function) of the relativistic
Coulomb system. Different from the former papers \cite{12,13} where the
relativistic Coulomb system is solved by the path integral with the KS
transformation and a beautiful perturbation technique, respectively. The way
presented here is more compact and suitable for arbitrary one dimensional
and spherical symmetry systems.

\section{DURU-KLEINERT METHOD FOR the RELATIVISTIC POTENTIAL PROBLEMS}

~~~~~Adding a vector potential ${\bf A(x)}$ to Kleinert's path integral for
a relativistic particle in a potential $V({\bf x})$ \cite{12}, we find that
the expression of the fixed-energy amplitude of a relativistic particle in
external static electromagnetic fields is given by \cite{14} 
\begin{equation}
G({\bf x}_{b},{\bf x}_{a};E)=\frac{i\hbar }{2mc}\int_{0}^{\infty }dL\int 
{\cal D}\rho (\lambda )\Phi \left[ \rho (\lambda )\right] \int {\cal D}%
^{D}x(\lambda )e^{-A_{E}\left[ {\bf x},{\bf x}^{\prime }\right] /\hbar }
\label{1}
\end{equation}
with the action 
\begin{equation}
A_{E}\left[ {\bf x},{\bf x}^{\prime }\right] =\int_{\lambda _{a}}^{\lambda
_{b}}d\lambda \left[ \frac{m}{2\rho \left( \lambda \right) }{\bf x}^{\prime
^{2}}\left( \lambda \right) -i(e/c){\bf A(x)\cdot x^{\prime }(}\lambda {\bf )%
}-\rho (\lambda )\frac{\left( E-V({\bf x})\right) ^{2}}{2mc^{2}}+\rho \left(
\lambda \right) \frac{mc^{2}}{2}\right] .  \label{2}
\end{equation}
where $L$ is defined as 
\begin{equation}
L=\int_{\lambda _{a}}^{\lambda _{b}}d\lambda \rho (\lambda ),  \label{3}
\end{equation}
in which $\rho (\lambda )$ is an arbitrary dimensionless fluctuating scale
variable, and $\Phi \lbrack \rho (\lambda )]$ is some convenient
gauge-fixing functional. The only condition on $\Phi \lbrack \rho (\lambda
)] $ is that [13-16] 
\begin{equation}
\int {\cal D}\rho (\lambda )\Phi \left[ \rho (\lambda )\right] =1.  \label{4}
\end{equation}
$\hbar /mc$ is the well-known Compton wave length of a particle of mass $m$, 
${\bf A(x)}$ is the vector potential, $V({\bf x})$ is the scalar potential, $%
E$ is the system energy, and ${\bf x}$ is the spatial part of the ($D+1$)
vector $\vec{x}=({\bf x},\tau )$. This path integral forms the basis for
studying relativistic potential problems.

The relativistic path integral of Eq. (\ref{1}) in the absence of the vector
potential ${\bf A(x)}$ has a more elegant representation providing the new
path integral solutions via well-known ones if the systems are in
two-dimensional Minkowski space or rotationally invariant systems in any
dimensions. By decomposing the Eq. (\ref{1}) into angular parts and take
DK-transformation, we get \cite{15,17,18} 
\begin{equation}
G({\bf {x}}_{b},{\bf {x}}_{a};E)=\frac{1}{(r_{b}r_{a})^{(D-1)/2}}%
\sum_{l=0}^{\infty }G_{l}^{{\rm DK}}(r_{b},r_{a};E)\sum_{{\bf \hat{m}}}Y_{l%
{\bf {\hat{m}}}}({\bf {\hat{x}}}_{b})Y_{l{\bf \hat{m}}}^{\ast }({\bf \hat{x}}%
_{a}),  \label{5}
\end{equation}
where superscript ${\rm DK}$ indicates that the system has been performed by
the DK-transformation, the functions $Y_{l{\bf \hat{m}}}(\hat{x})$ are the $D
$-dimensional hyperspherical harmonics and $G_{l}^{{\rm DK}}(r_{b},r_{a};E)$
is the purely radical transformed fixed-energy amplitude 
\begin{equation}
G_{l}^{{\rm DK}}(r_{b},r_{a};E)=\frac{\hbar i}{2mc}%
f_{b}^{1/4}f_{a}^{1/4}G(q_{b,}q_{a};{\cal E}).  \label{6}
\end{equation}
The amplitude $G(q_{b,}q_{a};{\cal E})$ of the fixed-pseudoenergy ${\cal E}$
is given by \cite{15,17,18} 
\begin{equation}
G(q_{b},q_{a};{\cal E})\equiv \int_{0}^{\infty }dS\int {\cal D}\rho (s)\Phi %
\left[ \rho (s)\right] \int {\cal D}q(s)e^{-A_{s}^{{\rm DK}}\left[ q,\dot{q}%
\right] /\hbar }  \label{7}
\end{equation}
with 
\[
A_{s}^{{\rm DK}}\left[ q,\dot{q}\right] =\int_{0}^{S}ds\left[ \frac{m}{2\rho
(s)}\dot{q}^{2}(s)+\rho (s)f(q(s))\right. \qquad \qquad \qquad \qquad 
\]
\begin{equation}
\left. \times \left( \frac{\hbar ^{2}}{2m}\frac{(l+D/2-1)^{2}-1/4}{%
r^{2}(q(s))}-\frac{\left[ E-V(r(q))\right] ^{2}}{2mc^{2}}+\frac{mc^{2}}{2}%
\right) +V{\rm _{eff}}(q(s))\right] ,  \label{8}
\end{equation}
where the effective potential $V_{eff}$ has the form  
\begin{equation}
V{\rm _{eff}}(q(s))=-\frac{\rho (s)\hbar ^{2}}{m}\left[ \frac{1}{4}\frac{%
h^{\prime \prime \prime }(q)}{h^{\prime }(q)}-\frac{3}{8}\left( \frac{%
h^{\prime \prime }(q)}{h^{\prime }(q)}\right) ^{2}\right]   \label{9}
\end{equation}
with $h^{\prime }(q)$ standing for the derivative $dh(q)/dq$ and the
transformation function $h(q)$ defined as $r=h(q)$ which is related to the
local space-time transformation function $f(r)$ \cite{15,17} by the
following equality 
\begin{equation}
h^{\prime 2}(q)=f(r).  \label{10}
\end{equation}

For the pure relativistic Coulomb system under consideration, the potential $%
V(r_{{\rm C}})=-e^{2}/r_{{\rm C}}${\rm \ }and the relativistic radial path
integral reads \cite{17,19} 
\begin{equation}
G_{l_{{\rm C}}}(r_{{\rm C}b},r_{{\rm C}a};E_{{\rm C}})=\frac{\hbar i}{2m_{%
{\rm C}}c}\int_{0}^{\infty }dL\int {\cal D}\rho (\lambda )\Phi \left[ \rho
(\lambda )\right] \int {\cal D}r_{{\rm C}}(\lambda )\exp \left\{ -\frac{1}{%
\hbar }A_{l}\left[ r_{{\rm C}},r_{{\rm C}}^{\prime }\right] \right\}
\label{11}
\end{equation}
with the action 
\[
A_{l}\left[ r_{{\rm C}},r_{{\rm C}}^{\prime }\right] 
\]
\begin{equation}
=\int_{\lambda _{a}}^{\lambda _{b}}d\lambda \left[ \frac{m_{{\rm C}}}{2\rho
\left( \lambda \right) }r_{{\rm C}}^{\prime ^{2}}\left( \lambda \right) +%
\frac{\rho \left( \lambda \right) \hbar ^{2}}{2m_{{\rm C}}}\frac{(l_{{\rm C}%
}+D_{{\rm C}}/2-1)^{2}-1/4}{r_{{\rm C}}^{2}}-\rho (\lambda )\frac{\left( E_{%
{\rm C}}+e^{2}/r_{{\rm C}}\right) ^{2}}{2m_{{\rm C}}c^{2}}+\rho \left(
\lambda \right) \frac{m_{{\rm C}}c^{2}}{2}\right] ,  \label{12}
\end{equation}
where the Roman subscript ${\rm C}$ specifies the Coulomb system. Let us
apply the DK-transformation to this relativistic system by taking the
following transformation variables 
\begin{equation}
\left\{ 
\begin{array}{l}
r_{{\rm C}}=h(x)=e^{x}, \\ 
h^{\prime 2}(x)=e^{2x}=f(r_{{\rm C}})=r_{{\rm C}}^{2},
\end{array}
\right.  \label{13}
\end{equation}
which maps the interval $r\in \left( 0,\infty \right) $ into $x\in \left(
-\infty ,\infty \right) $ and leads to the effective potential 
\begin{equation}
V{\rm _{eff}}(x(s))=\frac{\rho (s)\hbar ^{2}}{8m_{{\rm C}}}.  \label{14}
\end{equation}
It is without losing the generalization to take $\Phi \left[ \rho (\lambda )%
\right] =\delta \left[ \rho -1\right] $ and than the transformed
fixed-energy amplitude of Eq. (\ref{6}) turns into Morse potential system 
\[
G_{l_{{\rm C}}}^{{\rm DK}}(r_{{\rm C}b},r_{{\rm C}a};E_{{\rm C}})=\frac{%
\hbar i}{2m_{{\rm C}}c}f_{b}^{1/4}f_{a}^{1/4}G(x_{b,}x_{a};{\cal E}_{{\rm M}%
}) 
\]
\begin{equation}
=\frac{\hbar i}{2m_{{\rm C}}c}e^{x_{b}/2}e^{x_{a}/2}\int_{0}^{\infty }dS\int 
{\cal D}x(s)e^{-A^{{\rm DK}}\left[ x,\dot{x}\right] /\hbar }  \label{15}
\end{equation}
with the action 
\begin{equation}
A^{{\rm DK}}\left[ x,\dot{x}\right] =\int_{0}^{S}ds\left[ \frac{m_{{\rm C}}}{%
2}\dot{x}^{2}\left( s\right) +\frac{v^{2}\hbar ^{2}}{2m_{{\rm C}}}\left(
e^{2x}-2\alpha e^{x}\right) -{\cal E}_{{\rm M}}\right] .  \label{16}
\end{equation}
The parameters associated to the relativistic Coulomb system are given as 
\begin{equation}
\left\{ 
\begin{array}{l}
v=\frac{1}{\hbar c}\sqrt{m_{{\rm C}}^{2}c^{4}-E_{{\rm C}}^{2}} \\ 
\alpha =\frac{E_{{\rm C}}e^{2}}{m_{{\rm C}}^{2}c^{4}-E_{{\rm C}}^{2}} \\ 
{\cal E}_{{\rm M}}=-\frac{\hbar ^{2}}{2m_{{\rm C}}}\left[ (l_{{\rm C}}+D_{%
{\rm C}}/2-1)^{2}-\alpha ^{2}\right]
\end{array}
\right. ,  \label{17}
\end{equation}
where the notation $\alpha =e^{2}/\hbar c$ is the fine structure constant.
To go further, let's make a trivial transformation by taking 
\begin{equation}
\left\{ 
\begin{array}{l}
x=2x_{{\rm O}}, \\ 
m_{{\rm C}}=m_{{\rm O}}/4.
\end{array}
\right.  \label{18}
\end{equation}
This maps the Eq. (\ref{15}) into 
\begin{equation}
G_{l_{{\rm C}}}^{{\rm DK}}(r_{{\rm C}b},r_{{\rm C}a};E_{{\rm C}})=\frac{%
\hbar i}{2m_{{\rm C}}c}e^{x_{{\rm O}b}}e^{x_{{\rm O}a}}\frac{1}{2}%
\int_{0}^{\infty }dS\int {\cal D}x_{{\rm O}}(s)e^{-A^{{\rm DK}}\left[ x_{%
{\rm O}},\dot{x}_{{\rm O}}\right] /\hbar }  \label{19}
\end{equation}
with the transformed new action 
\begin{equation}
A^{{\rm DK}}\left[ x_{{\rm O}},\dot{x}_{{\rm O}}\right] =\int_{0}^{S}ds\left[
\frac{m_{{\rm O}}}{2}\dot{x}_{{\rm O}}^{2}\left( s\right) +\frac{2v^{2}\hbar
^{2}}{m_{{\rm O}}}\left( e^{4x_{{\rm O}}}-2\alpha e^{2x_{{\rm O}}}\right) -%
{\cal E}_{{\rm M}}\right] .  \label{20}
\end{equation}
The factor $1/2$ in Eq. (\ref{19}) accounts for the fact that the normalized
states are related by $\left| x\right\rangle =\left| x_{{\rm O}%
}\right\rangle /2.$ At this place, we can Apply the DK-transformation again
by taking the following transformation functions 
\begin{equation}
\left\{ 
\begin{array}{l}
x_{{\rm O}}=\ln z=h(z), \\ 
h^{\prime 2}(z)=1/z^{2}=f(x_{{\rm O}})=e^{-2x_{{\rm O}}},
\end{array}
\right.  \label{21}
\end{equation}
which maps the interval $x_{{\rm O}}\in \left( -\infty ,\infty \right) $
into $z\in \left( 0,\infty \right) $ and leads the effective potential $V%
{\rm _{eff}}$ $(z)$ to 
\begin{equation}
V{\rm _{eff}}(z)=-\frac{\hbar ^{2}}{8m_{{\rm O}}z^{2}}.  \label{22}
\end{equation}
The fixed-energy amplitude in Eq. (\ref{19}) becomes 
\begin{equation}
G_{l_{{\rm C}}}^{{\rm DK}}(r_{{\rm C}b},r_{{\rm C}a};E_{{\rm C}})=\frac{%
\hbar i}{2m_{{\rm C}}c}\frac{1}{2}z_{b}z_{a}\left\{ \frac{1}{\sqrt{z_{b}z_{a}%
}}\int_{0}^{\infty }dS^{\prime }\int_{0}^{\infty }{\cal D}z(\tau )e^{-A^{%
{\rm DK}}\left[ z,\dot{z}\right] /\hbar }\right\}  \label{23}
\end{equation}
with the action of the radial simple harmonic oscillator 
\begin{equation}
A^{{\rm DK}}\left[ z,\dot{z}\right] =\int_{0}^{S^{\prime }}d\tau \left[ 
\frac{m_{{\rm O}}}{2}z^{2}\left( \tau \right) +\frac{\hbar ^{2}}{2m_{{\rm O}}%
}\frac{(l_{{\rm O}}+D_{{\rm O}}/2-1)^{2}-1/4}{z^{2}}+\frac{m_{{\rm O}}\omega
^{2}z^{2}}{2}-{\cal E}_{{\rm O}}\right] .  \label{24}
\end{equation}
The parameter relations between the Eqs. (\ref{19}) and (\ref{24}) are given
as 
\begin{equation}
\left\{ 
\begin{array}{l}
\hbar ^{2}(l_{{\rm O}}+D_{{\rm O}}/2-1)^{2}=-2m_{{\rm O}}{\cal E}_{{\rm M}}
\\ 
m_{{\rm O}}\omega ^{2}/2=2v^{2}\hbar ^{2}/m_{{\rm O}} \\ 
{\cal E}_{{\rm O}}=4\alpha v^{2}\hbar ^{2}/m_{{\rm O}}
\end{array}
\right. .  \label{25}
\end{equation}
By inserting the relations in Eq. (\ref{17}) into these equality, we obtain
the parameter relations between the relativistic Coulomb and radial harmonic
oscillator 
\begin{equation}
\left\{ 
\begin{array}{l}
\mu _{{\rm O}}=2\sqrt{\mu _{{\rm C}}^{2}-\alpha ^{2}} \\ 
\omega =\sqrt{m_{{\rm C}}^{2}c^{4}-E_{{\rm C}}^{2}}/2m_{{\rm C}}c \\ 
{\cal E}_{{\rm O}}=E_{{\rm C}}e^{2}/m_{{\rm C}}c^{2}
\end{array}
\right. ,  \label{26}
\end{equation}
where for simplicity the quantity $(l_{{\rm O}}+D_{{\rm O}}/2-1)$ and $(l_{%
{\rm C}}+D_{{\rm C}}/2-1)$ have been defined as $\mu _{{\rm O}}$ and $\mu _{%
{\rm C}}$, respectively. With the well-known fixed-energy amplitude of a
radial harmonic oscillator 
\[
G_{l_{{\rm O}}}(z_{b},z_{a};{\cal E}_{{\rm O}})=-i\frac{1}{\omega }\frac{%
\Gamma ((1+\mu _{{\rm O}})/2-{\cal E}_{{\rm O}}/2\hbar \omega )}{\Gamma
(1+\mu _{{\rm O}})\sqrt{z_{b}z_{a}}} 
\]
\begin{equation}
\times W_{{\cal E}_{{\rm O}}/2\hbar \omega ,\mu _{{\rm O}}/2}((m_{{\rm O}%
}\omega /\hbar )z_{b}^{2})M_{{\cal E}_{{\rm O}}/2\hbar \omega ,\mu _{{\rm O}%
}/2}((m_{{\rm O}}\omega /\hbar )z_{a}^{2}),  \label{27}
\end{equation}
we finally obtain the exact radial fixed-energy amplitude of the
relativistic Coulomb system \cite{13} 
\[
G_{l_{{\rm C}}}(r_{{\rm C}b},r_{{\rm C}a};E_{{\rm C}})=\frac{m_{{\rm C}}c}{%
\sqrt{m_{{\rm C}}^{2}c^{4}-E_{{\rm C}}^{2}}}\qquad \qquad \qquad 
\]
\[
\times \frac{\Gamma \left( 1/2+\sqrt{(l_{{\rm C}}+D_{{\rm C}%
}/2-1)^{2}-\alpha ^{2}}-E_{{\rm C}}\alpha /\sqrt{m_{{\rm C}}^{2}c^{4}-E_{%
{\rm C}}^{2}}\right) }{\Gamma \left( 1+2\sqrt{(l_{{\rm C}}+D_{{\rm C}%
}/2-1)^{2}-\alpha ^{2}}\right) } 
\]
\[
\times W_{E_{{\rm C}}\alpha /\sqrt{m_{{\rm C}}^{2}c^{4}-E_{{\rm C}}^{2}},%
\sqrt{(l_{{\rm C}}+D_{{\rm C}}/2-1)^{2}-\alpha ^{2}}}\left( \frac{2}{\hbar c}%
\sqrt{m_{{\rm C}}^{2}c^{4}-E_{{\rm C}}^{2}}r_{{\rm C}b}\right) 
\]
\begin{equation}
\times M_{E_{{\rm C}}\alpha /\sqrt{m_{{\rm C}}^{2}c^{4}-E_{{\rm C}}^{2}},%
\sqrt{(l_{{\rm C}}+D_{{\rm C}}/2-1)^{2}-\alpha ^{2}}}\left( \frac{2}{\hbar c}%
\sqrt{m_{{\rm C}}^{2}c^{4}-E_{{\rm C}}^{2}}r_{{\rm C}a}\right) .  \label{28}
\end{equation}
This complete the discussion of Duru-Kleinert equivalence between the
relativistic Coulomb system and a radial harmonic oscillator.

\section{Concluding remarks}

In this paper, the DK-method is applied to the relativistic path integral.
As an interesting application, the fixed-energy amplitude of the
relativistic Coulomb system is solved by the DK-equivalence of the
relativistic Coulomb and a radial harmonic oscillator. Since the equivalence
is between the relativistic and non-relativistic physical problems, it may
provides us a more unified viewpoint of the different physical problems.
Different from the path integral approach \cite{12} and the perturbation
approach \cite{13}, the method presented in the paper just need to find the
appropriate transformation functions. Furthermore, all one dimensional and
any higher dimensional system with rotationally invariant systems are
suitable. We hope that the method presented here offers us a new way for
solving the relativistic potential problems.

\end{document}